\documentclass[12pt, executivepaper]{article}
\usepackage{times}
\usepackage[utf8]{inputenc}
\usepackage{seqsplit}
\usepackage[T1]{fontenc}
\usepackage{orcidlink}
\usepackage{amsmath, xfrac}
\usepackage{color}
\usepackage{amsfonts}
\usepackage{amsmath, amssymb}
\usepackage{graphicx}
\usepackage{setspace}

\newcommand{\RP}{{RP}}
\newcommand{\RPspec}[1]{{RP}^{\mathrm{spec}}_{\mathrm{#1}}}
\newcommand{\RPsh}{{RP}^{\mathrm{sh}}_{\mathrm{MET}}}

\usepackage{tikz}
\usetikzlibrary{arrows.meta, positioning}
\usepackage{enumitem}
\usepackage{geometry}
\usepackage{tikz}
\usetikzlibrary{arrows.meta,positioning,decorations.pathreplacing}

\usepackage{dcolumn, braket, graphicx, xcolor, bm, lipsum, cases, epigraph}
 \usepackage{mathptmx}
\usepackage{hyperref}
\hypersetup{colorlinks= true, allcolors=blue}
\usepackage{natbib}
\usepackage{tikz}
\usetikzlibrary{arrows.meta, positioning, decorations.pathreplacing}

\setcitestyle{authoryear,open={(},close={)}}

\title{\textbf{\mbox{Do Machine Learning Models Predict?}}}
\usepackage{orcidlink}      

\author{
  Michele Pizzochero\textsuperscript{1,}\thanks{E-mail: {mp2834@bath.ac.uk}}\textsuperscript{\hspace{0.6em}\orcidlink{0000-0003-3948-8202}}\hspace{0.5em}and\hspace{0.4em}Emanuele Ratti\textsuperscript{2,}\thanks{E-mail: {emanuele.ratti@bristol.ac.uk}}\textsuperscript{\hspace{0.6em}\orcidlink{0000-0003-1409-8240}}\\[0.8ex]
  \footnotesize\textsuperscript{1}\emph{Department of Physics \& Centre for Artificial Intelligence, University of Bath, United Kingdom}\\[-0.3ex]
  \footnotesize\textsuperscript{2}\emph{Department of Philosophy, University of Bristol, United Kingdom}
}

\date{}
\begin{document}
\begin{singlespace}
\maketitle 
\end{singlespace}

\begin{abstract}
Machine learning (ML) models are said to predict. Philosophers draw a line between accommodation and prediction, depending on whether or not the evidence a theory succeeds on was used in constructing it. Does this distinction survive when a theory is replaced by an ML model? We argue so, yet the dividing line becomes fuzzy. Whether an output of an ML model counts as an accommodation or as a prediction hinges upon the relationship between the patterns the input instantiate and those on which the model was selected through its training. Once this relation is taken into account, most successful outputs of ML models, we suggest, bear a strong accommodation character. We further trace the consequences of our account for confirmation, novelty, and scientific realism.
\end{abstract}
\begin{singlespace}
\vskip -1.5cm
\tableofcontents
\end{singlespace}
\newpage

\section{How Predictive Are Machine Learning Predictions?}
A convolutional neural network trained on lymph node images is presented with a patch of tissue. It outputs: \textsc{metastasis}. In Machine Learning (ML) speak, the network has made a \textit{prediction}. The term is ubiquitous: AlphaFold predicts three-dimensional protein structures from their amino-acid sequences; GraphCast predicts the weather for the following week; and GPT predicts the next word of a text. An ML model shown an input identical or very similar to one it was trained on makes a prediction, just as much as a model shown an input it has never encountered. Nothing in the ML usage of the term requires the output to concern, e.g., the future, or the unseen, or to be risky.

Philosophers reserve the term prediction for something more demanding. A prediction is contrasted with an \textit{accommodation}. Roughly, a theory or model is said to accommodate evidence it was built to fit, and to predict evidence it was not. Whether this difference carries epistemic weight has been debated since at least the nineteenth century, when \citet{Whewell:1840} and \citet{Mill:1843} disagreed over the merits of prediction and accommodation. \textit{Predictivists} hold that predicted evidence confirms a theory, either directly or indirectly, more strongly than merely accommodated evidence. This thesis has been drawn on in debates ranging from the credentials of confirmed theories to scientific realism. The following line of thought might be tempting: ML models are said to predict; predictivism says prediction matters; predictivism applies to ML models.\footnote{ML outputs are termed predictions because ML models are predictors in the statistical sense: they estimate an output from an input. We do not propose to change this usage, and our question is not merely semantic. Science prizes predictions as a route to confirmation, so whether ML outputs are predictions in the philosophers’ sense bears on what ML-driven discoveries can establish.
} Not so fast, we argue.

In this article, we ask whether, and to what degree, the outputs of ML models should be regarded as predictions in the philosophers’ sense. We claim that whether an output of an ML model counts as a prediction or as an accommodation hinges on the relation between the patterns the input instantiate and those on which the model was selected through its training. Once this relationship is taken into account, most successful outputs of ML models, we suggest, carry a strong accommodation character.

We proceed as follows. In Section \ref{Section-2}, we reformulate the notions of prediction and accommodation, hitherto developed for scientific theories, for ML models. In Section \ref{Section-3}, we defend the view that ML outputs that matter most, those on inputs unlike the training data, are neither pure predictions nor pure accommodations. In Section \ref{Section-4}, we fuzzify the two notions and introduce the prediction score, a metric that places an ML output on the resulting prediction-accommodation continuum, further casting doubt on the ability of ML models to confidently produce outputs with a dominant prediction character.  In Section \ref{Section-5}, we conclude by highlighting the consequences of our account for confirmation, novelty, and scientific realism in the context of ML. Throughout, we illustrate our arguments by training a simple convolutional neural network for detecting cancer metastases, in the hope of engaging philosophers and scientists alike.

\section{Prediction and Accommodation, Revisited}
\label{Section-2}

\paragraph{Prediction and accommodation for scientific theories.}  What counts as a prediction, as opposed to an accommodation? In explaining the basic intuition behind predictivism, \citet{Barnes:2009} appeals to the notion of novelty, that is, ``when $E$ is evidence for $T$, $E$ supports $T$ more strongly when it is a \textit{novel} confirmation'' (p.\ 1; emphasis added). Successfully predicting a novel phenomenon is thus regarded as a virtue of a given \textit{T}. When Einstein’s general relativity predicted the gravitational deflection of starlight by the Sun, later confirmed by Eddington's expedition, the effect was dramatic. Had the theory merely been finely tuned to reproduce such deflection, the same agreement would have counted for less.

Central to this debate is what makes a confirmation novel. The first answer, offered by  \citet{Lakatos:1970}, is temporal: $E$ is novel if it was not known when $T$ was constructed. This drew strong criticism, specifically because when $E$ was discovered seems to have no bearing on the extent to which $E$ supports $T$. Insofar as a logic of confirmation ought to be concerned with the  relationship between theory and evidence, the temporal dimension should be irrelevant. As an alternative to this proposal, some philosophers have put forward heuristic accounts. These relocate `novelty' from the history of $E$ to the construction of $T$. According to \citet{1dc5ef8d-d2b2-347c-90e1-eee27703f1c3}, $T$ accommodates $E$ if it was built to fit $E$, and predicts $E$ if it was not. The criterion thus shifts from when $E$ became known to how $T$ was constructed: accommodated evidence is evidence that was used in constructing $T$. The resulting difference is epistemic and not merely psychological, as it can be used to adjudicate between an \emph{ad hoc} rescue of a theory and genuine (dis)confirmation, while also bearing on  the impact of overfitting in the assessment of theories, as \citet{8baa96f9-5f2e-3fe1-a31b-c8e72f8b5be8} point out.

The starting point of our discussion is \citet{10.1093/mind/112.448.653}'s distinction between prediction and accommodation, which is representative of heuristic definitions. As we read it, it is tailored to win acceptance among philosophers who disagree about predictivism, and it commits to nothing controversial. Let $T$ be a theory that entails a proposition $P$, which is discovered to be true. The $P$ that matters here is relative to an empirical observation concerning the datum $D$. \citet[p.~655]{10.1093/mind/112.448.653}
defines:

\begin{quote}
\emph{Accommodation}. $T$ accommodates $D$ iff $T$ entails $D$, $D$ is true, and $T$ was constructed exactly to entail $D$.

\emph{Prediction}. $T$ predicts $D$ iff $T$ entails $D$, $D$ is true, and $T$ was not constructed to entail $D$.
\end{quote}
The heuristic character of these definitions lies in their concern with the origin of $T$ and its causal relationship with $D$, and not with the point in time in which $D$  became known. 

Our plan is to use White's definitions in the context of ML. However, before we can ask whether an output of an ML model is a prediction or accommodation in White's sense, these definitions must be translated from theories to ML models. To that end, three points need to be addressed. The first pertains to the fact that the distinction is presented for scientific theories, and an ML model is not obviously one. Second, White's definitions require $D$ to be true, whereas an ML output is generally viewed as only correct or incorrect. Third, while $T$ entails $D$, an ML model does not  entail anything, since it learns by induction and speaks in probabilities. We take these issues in turn, and end by offering a definition of accommodation and prediction applicable to ML models.

\paragraph{Prediction and accommodation for ML models.}  We adopt the conception of ML models articulated by \citet{Rattietal:2026} and \citet{10.1007/978-3-031-09153-7_7}. On this conception,  an ML model is a computational artifact. As such, a full picture of what an ML model is \emph{qua} computational artifact requires, following \citet{RN686}, a description in terms of levels of abstraction. The same ML model is at once an abstract mathematical object, an algorithm, a piece of software, and an electronic computation running on physical hardware. These views do not exclude one another, yet some questions are better settled at one level than at another. There are five levels of abstraction underlying a computational artifact:

\begin{enumerate}
\item Functional Specification Level (FSL): the mathematical specification of the behavior of an artifact, in terms of the function it computes (e.g., for a computer simulation, a transition function taking the state of a physical system at time $t$ to its state at the next timestep $t$+$\Delta t$);
\item Design Specification Level (DSL): the abstract procedure for computing the FSL function (e.g., for a simulation, a set of equations);
\item Algorithm Design Level (ADL): the structured procedure for computing the DSL specification (e.g., for a simulation, the numerical methods used to solve the set of equations);
\item Algorithm Implementation Level (AIL): the translation of the ADL procedure into a programming language (e.g., a Python program);
\item Algorithm Execution Level (AEL): the physical execution of the program on hardware.
\end{enumerate}

An ML model can be described at the FSL, DSL, and ADL.\footnote{The remaining AIL and AEL do not differ between conventional computer simulations and ML models.} 
At FSL, an ML model is a probability distribution selected from a family of probability distributions. For classic discriminative models, the family is $\{p_\theta(Y\mid X):\theta\in\Theta\}$, where $X$ is the input space, $Y$ the output space, and $\theta$ the specific set of parameters chosen from all the possible parameter configurations, $\Theta$.  For generative models, this reads $\{p_\theta(X) : \theta \in \Theta\}$.
Training determines the numerical values of parameters by minimizing a loss
function or maximizing a likelihood, thus selecting one member of the family,
which is the actual ML model at FSL. At DSL, an ML model is specified by how
the main distribution factorizes into simpler conditional distributions, which
introduces components such as latent variables and conditional dependencies.
At ADL, an ML model is specified as an architecture blueprint, which spells
out the DSL specification through, e.g., layers, convolutional kernels,
pooling operations, attention heads, and so on. In other words, ADL defines
``what kinds of operations are available for transforming inputs into outputs,
and what constraints are imposed on those operations'' \citep{Rattietal:2026}.

With this framework in hand, we next turn to what \citet{10.1093/mind/112.448.653} means by  `theory.' A theory, he writes, is  ``a \emph{proposition} which cannot be molded into shape to fit the data. [...] A better metaphor for the process of theorizing is that of \emph{selecting} a theory off the platonic library shelf'' (p.\ 661; emphasis added). The idea that $T$ is selected, rather than molded, resonates well with ML models, since selecting is exactly what training does. The hyperparameters and architecture typically fix a space of possible parameter configurations (i.e., the weights). At FSL, a model is the probability distribution picked out by one configuration of weights within the space delimited at ADL. In other words, training is the process of selecting a configuration of weight-parameters (i.e., a single model) from among the set of potential parameters that the architecture permits at ADL. The set of all such possible configurations plays the role of White’s platonic library, and the trained model, at FSL, is the one selected off the library shelf. White’s metaphor applies to ML models almost without adjustment.

For an ML model, what would count as a prediction or an accommodation is its output. The nature of the output depends on the specific task, be that a label (as in the classification of a neural network), a number (as in the temperature forecast by a weather model), or a word (as in the next token of a large language model). In each case, the output is a value that the ML model assigns to an input. White holds that $T$ entails $D$ and that $D$ is true. We have identified $T$ with an ML model, so $D$ is the datum that the ML model gets right or wrong: a given output assigned to a given input. Is the output ‘true’ in White’s sense? We could force this reading. Labeling a patch of metastatic tissue as \textsc{metastasis} says something true about the patch, namely that it shows metastatic tissue. Yet, it is perhaps safer to speak of \textit{correctness} instead of truth. An output is correct when it meets a standard of accuracy that is established independently of the model (that is, the ground truth), such as a pathologist's expert annotation of a laboratory immunohistochemical test. 
Correctness thus plays in our analysis the role that truth plays in White’s. This choice is deliberate: what is at stake is the discovery potential of ML in science, so we ask whether models predict or accommodate when their outputs are correct. In ML parlance, a model ‘predicts' whenever it returns an output for a given input. Since any such output can typically be checked against a ground truth, it can be right or wrong, and it is only for the right ones that the philosophical distinction between prediction and accommodation becomes relevant.

Finally, what about entailment? Is there any sense of entailment at play in ML models? \citet{10.7551/mitpress/5876.001.0001} point out that, through training, ML models learn by induction. Once ADL is fixed, training selects the parameters once and for all, and such parameters, together with the fixed operations of the architecture, settle what comes out when an input is presented. The output for a given input is then determined in a deductive-like fashion. Just as a theory $T$, given some particular initial conditions and auxiliary assumptions, entails a particular datum $D$, a trained ML model $M$ (e.g., a convolutional neural network), given an input $x$ (e.g., a patch of lymph node tissue), entails a particular output $y$ (e.g., \textsc{metastasis}). So we can say that the ML model, applied to $x$, entails $y$. This gives us the contrast we need. If the ML model was not selected to entail $y$ for $x$, then assigning $y$ to $x$ is a prediction. If it was, it is an accommodation.

With these three points settled, we arrive at the definitions of accommodation and prediction for ML models:

\begin{quote}
\emph{Accommodation}. An ML model $M$, assigning an output $y$ to an input $x$ is an accommodation iff applying $M$ to $x$ entails $y$ for $x$, $y$ is correct for $x$, and $M$ was selected to entail $y$ for that $x$.

\emph{Prediction}. An ML model $M$, assigning an output $y$ to an input $x$ is a prediction iff applying $M$ to $x$ entails $y$ for $x$, $y$ is correct for $x$, and $M$ was not selected to entail $y$ for that $x$.

\end{quote}

We emphasize that these definitions are general and apply equally well to other ML outputs, such as numerical values, clusters, or tokens. In all cases, the ML model, characterized by its own FSL, DSL, and ADL, entails an output in response to an input, with the correctness of the output being established by a model-independent ground truth, with a tolerance suitable to the task. Whenever the output is a numerical value, such as in a regression model, the ground truth may be an acceptable numerical interval around a measured value; whenever the output is an assignment to a cluster, such as in an unsupervised model, the ground truth may be membership in a known category; and whenever the output is a sequence of words, such as in large language models, the ground truth may be a range of acceptable continuations, any of which may count as correct. Armed with these definitions, we can now ask whether ML outputs fall under predictions or under accommodations.

\section{Neither Pure Prediction Nor Pure Accommodation}
\label{Section-3}

\paragraph{A simple experiment.} To make our discussion concrete, we work through an illustration. We train a convolutional neural network to detect breast cancer metastases in lymph node tissue. We rely on ResNet18 trained from scratch (that is, from random initial weights), so that its training data are the sole source of what it learns. Presented with a small patch of tissue, its task is to decide whether or not the tissue is metastatic.

\begin{figure*}[t]
  \centering
  \includegraphics[width=\textwidth]{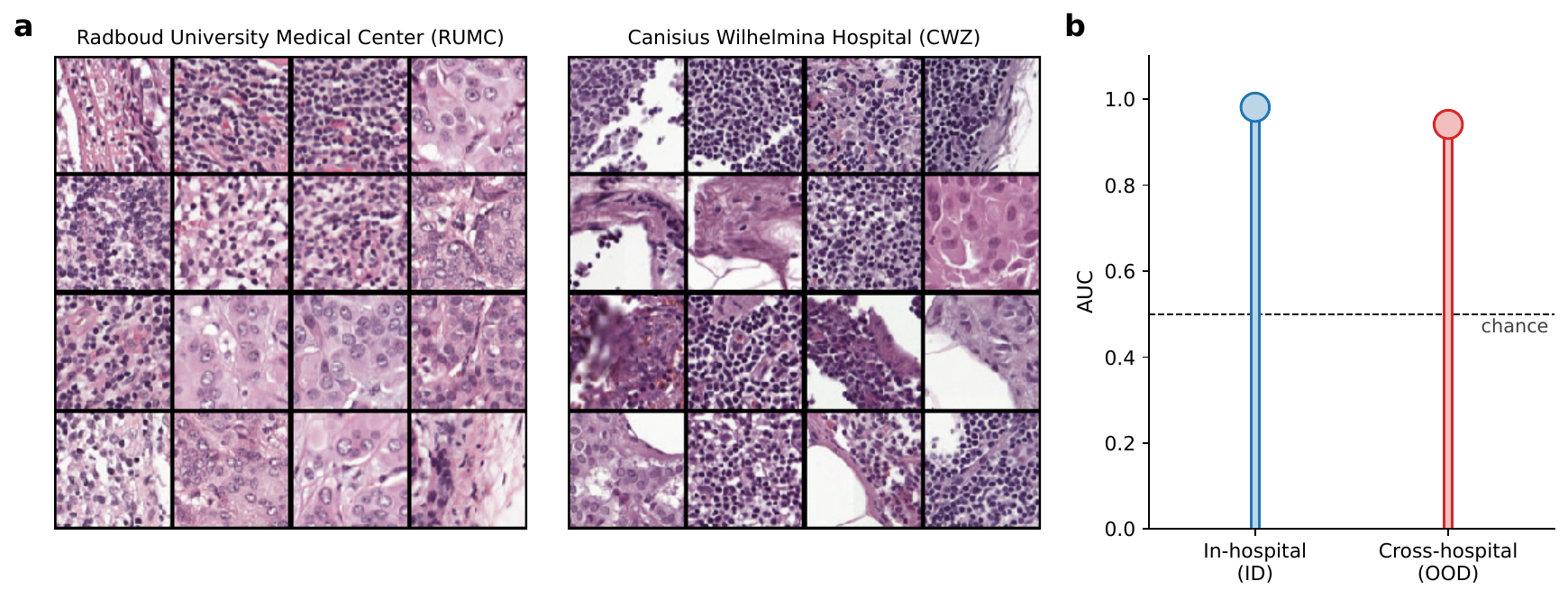}
  \caption{
    \textbf{a}, Random patches (96$\times$96 pixels; 4$\times$4 per hospital) cropped from
    hematoxylin-and-eosin-stained whole-slide images of sentinel lymph node sections from breast cancer patients,
    for the two hospitals considered, RUMC and CWZ.
    \textbf{b}, AUC of ResNet18 models trained from scratch on RUMC over five seeds, i.e., 0.98 $\pm$ 0.02 for in-hospital (ID), as obtained on two held-out RUMC slides, and 0.94 $\pm$ 0.01 for cross-hospital (OOD), as obtained on all 34,904 patches of CWZ. The dashed line marks
    chance level (AUC\,=\,0.5).}
  \label{Figure-1}
\end{figure*}

We consider a collection of patches cropped from digitized, hematoxylin-and-eosin-stained sections of sentinel lymph nodes from breast cancer patients. Each patch is labeled metastatic if the central region overlaps a metastasis that pathologists annotated on the slide, and non-metastatic otherwise. These pathologists' annotations provide the independent standard of correctness. The patches are from two Dutch hospitals, the Radboud University Medical Center (RUMC) and Canisius Wilhelmina Hospital (CWZ). Each hospital contributes ten fully annotated slides, from which thousands of patches are cropped. The patches are perfectly balanced between the two classes. RUMC and CWZ prepared and stained their slides in different ways. In Figure \ref{Figure-1}(a), we show sixteen representative patches for each of the two hospitals. Following training,\footnote{We trained ResNet18 \citep{he:2016} from random initial weights, with the Adam optimizer (learning rate $10^{-3}$, cosine decay), batch size 128 and 20 epochs, using random flips and brightness and color jitter as augmentation. Each model was trained on up to 32,000 patches drawn at random from one hospital, with two of its ten slides held out; the model after the final epoch is used, and nothing was selected using the held-out slides or the test hospital. We trained five models per direction, with different random seeds (the seed sets the initial weights, the patch sample, the held-out slides and the order of training), and report means and standard deviations over seeds. Performance is measured on all patches of the other hospital. The data is from CAMELYON17-WILDS \citep{Bandi:2019, Koh:2021}.} we evaluate the model performance through the Area Under the Curve (AUC).\footnote{AUC is a metric that quantifies how well our binary classification model can distinguish between the two metastatic and non-metastatic classes, i.e., the probability that a randomly chosen metastatic patch receives a higher probability than a randomly chosen non-metastatic one. Specifically, AUC = 1 denotes a perfect separation of the two classes, AUC = 0.5 denotes chance, while AUC = 0 denotes complete inversion of the two classes.}

We train the model on the patches from RUMC, $x\textsubscript{RUMC}$. We refer to the resulting model as $M\textsubscript{RUMC}$. At FSL, a model like $M\textsubscript{RUMC}$ is a conditional probability distribution $p_\theta(y\mid $x\textsubscript{RUMC}$)$, where $x\textsubscript{RUMC}$ is a patch from RUMC, and $y$ is the diagnostic label, such as \textsc{metastasis}, that is selected among a large family of possible distributions, where the boundaries are set by ADL. We then determine the performance of $M\textsubscript{RUMC}$ on two held-out slides (4,264-14,670 patches, depending on seed initialization) from RUMC, which the model never saw in training (in-hospital performance). These inputs, though unseen, are in-distribution (ID): they were prepared and stained like the training data belonging to the same hospital. As shown in Figure  \ref{Figure-1}b, we  obtain AUC$\textsubscript{RUMC} = 0.98 \pm 0.02 $. This nearly-perfect  AUC suggests that $M\textsubscript{RUMC}$ captures two kinds of regularity: some that are specific to RUMC, involving its own procedures in the slide preparation, and others that are related to a phenomenon in the real world that pathologists refer to as ‘metastasis.’

We next present to $M\textsubscript{RUMC}$, trained exclusively on $x\textsubscript{RUMC}$, all  34,904 patches from the hospital CWZ, $x\textsubscript{CWZ},$ which the model never saw in training (cross-hospital performance). As compared to the previous case, there is now a shift from in-distribution to out-of-distribution (OOD) data, because the two hospitals prepared and stained their slides differently, that is, there is a distribution shift as a result of different procedures leading to partially different data-generating processes. As shown in Figure  \ref{Figure-1}(b), we obtain AUC$\textsubscript{CWZ} = 0.94 \pm 0.01$. Because AUC$\textsubscript{CWZ}$ is close to AUC$\textsubscript{RUMC}$, with a gap of only $0.04 \pm 0.02$, we conclude that $M\textsubscript{RUMC}$ generalizes well, which suggests that it has captured regularities that carry over from one hospital to the other. Is a correct output on an input that $M\textsubscript{RUMC}$ has not seen a prediction? It seems straightforward: $M\textsubscript{RUMC}$ was not selected to entail this output for this very input, so the output appears to fall under the definition of prediction. We argue otherwise.

\paragraph{Real patterns as a criterion.} We connect our characterization of an ML model to \citet{946f4cb6-15e5-3a7b-921e-53dc0743f655}'s notion of real patterns. Dennett presents six images consisting of black and white dots, which he calls `frames'. No two frames are exactly alike, yet they can all be said to display a similar pattern consisting of five equally spaced black bars (a pattern that Dennett terms `bar code'). How does one establish whether a pattern is \textit{really} there? According to Dennett, one can rephrase this question in terms of the task of transmitting information about one of the frames. One way to do this would be to transmit the exact bit map, that is, to describe the frame in terms of the position of each of its black and white dots. In this case, \citet{946f4cb6-15e5-3a7b-921e-53dc0743f655} argues, for ``a black-and-white frame of 900 dots (or pixels, as they are called), the transmission requires 900 bits'' (p.\ 32). But there might be a more efficient way of describing a given frame. One can, for instance, describe some of the frames in terms of number and size of the bar codes. If a frame can be described in a way that is more efficient than the \textit{seriatim} description, then it is compressible, and the pattern is real. \citet{946f4cb6-15e5-3a7b-921e-53dc0743f655} concludes that a ``pattern exists in some data, is real, if there is a description of the data that is more efficient than the bit map'' (p.~34). If, on the other hand, ``nothing shorter than the verbatim bit map will preserve the series'' (p.~33), then a frame is random and incompressible. 

Now, take a frame to be a chunk of data, such as an individual input (e.g., a lymph node patch), whether inside or outside the training set, and suppose we want to transmit the information it contains. If the information can be described more efficiently than  reporting all the values of every pixel (that is, Dennett's \textit{seriatim} description), then there is a pattern in it. Can then the task of selecting an ML model be conceptualized as the search for real patterns? Yes. Consider an ML model, say, a convolutional neural network such as $M\textsubscript{RUMC}$, more closely. At FSL, the model is a conditional probability distribution $\{p_\theta(Y\mid X):\theta\in\Theta\}$, where $X$ is the input space, $Y$ the output space, and $\theta$ the specific set of parameters characterizing the distribution.  $\theta$ is selected via optimization by, e.g., minimizing a loss. Selecting an ML model can be seen as a form of compression, because what an ML model (understood at FSL) provides are `compact' rules \citep{10.7551/mitpress/5876.001.0001} for how configurations of the features of the inputs map into the output space. This amounts to saying that the model is a more efficient description of the mapping than the list of all the details of the inputs and their labels. A real pattern can thus be expressed as a (conditional) distribution that yields a compressed description of the frame with respect to an output space. If this is right, training an ML model can be understood as the search for real patterns in the data.

In the case of patches from RUMC, the near-perfect AUC$\textsubscript{RUMC} = 0.98 \pm 0.02 $  suggests that $M\textsubscript{RUMC}$ has identified patterns characteristic of how RUMC prepares and stains its slides as well as patterns relevant to metastasis (or, at least, to the part of the image that a pathologist would call `metastasis'). More generally, the output that an ML model produces depends on the real patterns that efficiently describe the input. An ML model entails $y$ for an input $x$ in virtue of $x$ instantiating certain real patterns. This suggests a criterion. We claim that model selection operates on real patterns, not on individual inputs. A model is thus selected to entail $y$ for $x$ to the extent that $x$ instantiates real patterns on which it was selected.

Let $\RP\textsubscript{RUMC}$ be the real patterns that efficiently describe the inputs $x_{\mathrm{RUMC}}$, and $\RP\textsubscript{CWZ}$ those that describe $x_{\mathrm{CWZ}}$. Each has a part tied to the hospital's own measurement apparatus and procedures, $\RPspec{RUMC}$ and $\RPspec{CWZ}$, reflecting how each hospital prepared and stained its slides, and a part that the two share, $\RPsh$, which is the set of patterns that distinguish metastatic from non-metastatic tissue. The cross-hospital model performance, AUC$\textsubscript{CWZ} = 0.94 \pm 0.01$,  motivates the latter part: $M\textsubscript{RUMC}$ generalizes from $x_{\mathrm{RUMC}}$ to $x_{\mathrm{CWZ}}$, as the small generalization gap ($0.04 \pm 0.02$) suggests, presumably because at least some of the patterns identified in the training data are also present in the unseen inputs, despite their having specific patterns of their own\footnote{Our argument does not presuppose any realist commitment. The `invariance' in question is across datasets, and nothing in what follows depends on how it is interpreted. One might nonetheless be tempted towards realism by the fact that a model that generalizes across more and more datasets is highly projectible. What is invariant could then be read as causal, and so as an instance of modal structure, as \citet{arjovsky2019invariant} do. The connection between modal structure and ML models has been noted before by, e.g., \cite{Rattietal:2026} and \cite{RN688}. Whether such a reading is warranted is contentious. We remain agnostic about it, and nothing in our argument depends on it.}. For simplicity, then,
\begin{align}
  \RP\textsubscript{RUMC} &= \RPspec{RUMC} \cup \RPsh, \\
  \RP\textsubscript{CWZ}  &= \RPspec{CWZ}  \cup \RPsh.
\end{align}

We can now ask whether a correct output of $M\textsubscript{RUMC}$  is a prediction or an accommodation, first for ID inputs, which follow the distribution of the training data, and then for OOD inputs, which do not. The decomposition above says what is at stake in each case: for the former, all of what an input instantiates was selected on, and for the latter, only part of it.

\paragraph{In-distribution (ID) inputs.} Show the ML model one of the very same inputs it was trained on (e.g., one of the RUMC patches in the training set to $M\textsubscript{RUMC}$), and the model answers correctly. This is an accommodation, plainly. The ML model was selected to entail that very output for that very input, so the answer was in the model before the question was asked. That is the trivial, and not very interesting, case. Consider another case. Show the ML model an input that it has not seen, but that comes from the same distribution, such as a held-out RUMC patch to $M\textsubscript{RUMC}$. In the idealized limit, unseen inputs from RUMC are generated exactly like the training inputs. This never happens in practice, since some difference remains, however standardized the procedure, and the held-out patches only approximate the limit, but let us take it. Such an input instantiates $\RP\textsubscript{RUMC}$, that is, $\RPspec{RUMC}$ \emph{and} $\RPsh$, with $M\textsubscript{RUMC}$ being selected on exactly these patterns. The input did not constrain the selection in the sense of being among the training inputs. The real patterns it instantiates, however, did. By our criterion, $M\textsubscript{RUMC}$ was selected to entail the output for this input, and a correct output on it is an accommodation. For ID inputs, \emph{whether seen or unseen}, the  output is not a prediction. Quite the opposite: because such inputs share all of their real patterns with the training data on which the model was selected, the output is a pure accommodation.

\paragraph{Out-of-distribution (OOD) inputs.} Take now, as an input, a CWZ patch for which $M\textsubscript{RUMC}$ entails the correct label. Here the claim that the output is a prediction may look even more secure: $M\textsubscript{RUMC}$ was not selected to entail it for this input, no CWZ patch was included in training, and the input comes from a shifted distribution. We ask the same questions as before. The input instantiates $\RP\textsubscript{CWZ}$, that is, $\RPspec{CWZ}$ and $\RPsh$. $M\textsubscript{RUMC}$ was selected on $\RPsh$ but not on $\RPspec{CWZ}$. The output is therefore not a pure prediction: the success of $M\textsubscript{RUMC}$ rests, at least in part, on the distribution-invariant patterns that bear on the presence or absence of metastasis, $\RPsh$, which were part of what $M\textsubscript{RUMC}$ was selected to account for.  But it is not a pure accommodation either. The input also instantiates $\RPspec{CWZ}$, on which $M\textsubscript{RUMC}$ was not selected, so part of what the input is lies outside what the selection constrained.

In contrast with the pervasive ML speak, we claim that an ML model not having seen a particular input does not, by itself, make a correct output on it a prediction. What matters to the distinction between prediction and accommodation is the relationship between the real patterns that the input instantiates and those on which the model was selected. ID outputs are pure accommodations, as they share \emph{all} their real patterns with the training data. OOD outputs are neither pure accommodations nor pure predictions, as they share only \emph{some} of their real patterns with the training data. Consider each extreme in turn. For an OOD output to be a pure accommodation, its input would have to share all its real patterns with the training data. But this is not possible, since an input that did so would no longer be OOD: it would be ID. For an OOD output to be a pure prediction, its input would have to share none of its real patterns with the training data. But then nothing in the model selection would map the output to the input, and a correct output would be a stroke of luck, not a success that can  reliably be ascribed to the model. We therefore conclude that OOD outputs are mixed cases: they carry something of accommodation and something of prediction, yet cannot be neatly reduced to either.

\section{Prediction and Accommodation, Fuzzified}
\label{Section-4}

\paragraph{Prediction and accommodation on a continuum.} We settled on a criterion to determine whether an OOD input is a prediction or an accommodation: an output carries an accommodation  character to the extent that its input instantiates real patterns on which the model was selected, and a prediction character to the extent that it does not. Building on this, we propose a fuzzy conception of prediction and accommodation. On this conception, an output is not simply a pure prediction or a pure accommodation but is one or the other to a degree, in the same way that membership in a fuzzy category comes in degrees. The two limits are idealizations that real outputs will never reach. On the one hand, pure accommodation would require an input that instantiates \emph{only} the real patterns on which the model was selected. On the other hand, pure prediction would require the input sharing \emph{none} of its real patterns with those on which the model was selected. Between these limits lies a continuum, and the position of an output on it depends on the extent to which it occurs with, or without, the real patterns on which the model was selected.\footnote{One may ask whether this fuzzy conception proves too much, if extended beyond ML models. General relativity's prediction of the bending of starlight, for example, depends on gravitational structure already present in data that was available to Einstein, so the theory too might carry an accommodation character, and the distinction would then dissolve in science generally. It does not, and the reason is a difference in what constrains selection. A theory is selected under two types of constraint: the data it is fitted to, if any, and principles that do not originate from those data. GR was selected under domain-specific principles of the second kind, such as the equivalence principle and general covariance, so part of what it entails about starlight is secured by them and not by the data. That part is prediction-like even on our account. In ML models, by contrast, the constraints are generic and domain-agnostic (see \cite{Rattietal:2026}, Chapter~2), so what an ML model entails about a new input is secured almost entirely by the patterns extracted from its training data.}

\begin{figure*}[t]
  \centering
  \includegraphics[width=\textwidth]{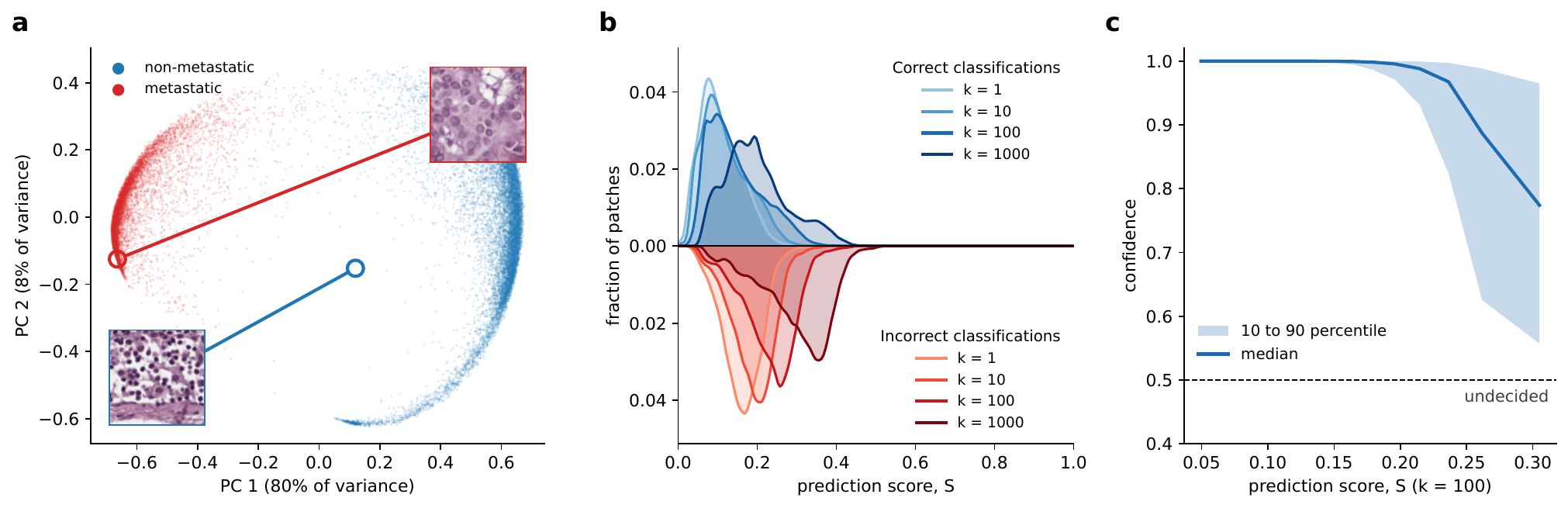}
  \caption{
    \textbf{a}, RUMC training patches in the first two principal components of the representation space (i.e., the 512-dimensional output of the network’s last pooling layer, normalized to unit length). Blue and red dots denote non-metastatic and metastatic patches, respectively. Circled, with their images, are two representative, correctly classified
    CWZ patches with the highest (top right inset) and lowest (bottom left inset) accommodation character.
    \textbf{b}, Distribution of the prediction score $S$ for correctly (blue) and incorrectly (red) classified CWZ patches for nearest-neighbor numbers $k = 1, 10, 100, 1000$.
    \textbf{c}, Model confidence (i.e., probability of the predicted class) for correctly classified
    patches: median (line) and 10th to 90th percentile (shading) in 20 equal-count bins of $S$ for $k = 100$. The dashed line marks confidence $0.5$, at which the model assigns equal probability to both metastatic and non-metastatic classes. All panels use one model.} 
  \label{Figure-2}
\end{figure*}

Real patterns cannot be inspected directly, though, so how is one going to establish where an output lies in the continuum? The model gives us a window onto them: its own representation space, that is, the space in which the model encodes each input, such that inputs that the network treats alike lie close together. We illustrate this by returning to $M\textsubscript{RUMC}$. In Figure~\ref{Figure-2}a, we show the 22,546 RUMC training patches, projected from the 512-dimensional representation space onto its first two principal components. They form two dense arcs, corresponding to the metastatic and non-metastatic classes. These are the regions where the patterns on which $M\textsubscript{RUMC}$ was selected are heavily represented. Any new OOD input is mapped by $M\textsubscript{RUMC}$ into the same space, and where it lands reveals how much its patterns were already represented in the selection. An input that lands in a densely populated region resembles many training inputs: $M\textsubscript{RUMC}$'s output on it rests on patterns that constrained the selection many times over, and is accommodation-like. An input that lands in a sparsely populated region, or in the empty gap between the arcs, resembles few or none of them: the selection hardly constrained it, and the output is prediction-like. This accommodation-like and prediction-like contrast is shown  in the insets of  Figure~\ref{Figure-2}a for two CWZ patches correctly classified by $M\textsubscript{RUMC}$. One lies on the arc of metastatic training patches, among many neighbors. The other lies in the empty interior between the two arcs, far from all of them. In other words, an OOD input acquires a stronger prediction character the farther it falls from the heavily represented regions of the model’s representation space. We now make the idea more quantitative.

\paragraph{The prediction score.} Let $z(x) \in \mathbb{R}^d$ be the representation of an OOD input $x$ in the last pooling layer of the network, normalized to unit length. The final layer of the network is a linear read-out of these features, so the output of the model for $x$ is fixed by its representation, and distances in this space are distances between inputs as the model sees them. For each  $x$, we define the \emph{prediction score}, $S$, as
\begin{equation}
  S \;=\; \frac{D_k(x)}{\sqrt{2}}, \qquad
  D_k(x) \;=\; \frac{1}{k}\sum_{j=1}^{k} \bigl\lVert z(x) - z\bigl(x^{(j)}\bigr) \bigr\rVert_2 ,
\end{equation}
where $x^{(1)},\dots,x^{(k)}$ are the $k$ training inputs whose representations are nearest to $z(x)$. In words, $S$ is the mean distance from an input to its $k$ nearest training inputs, rescaled to lie between $0$ and $1$. The factor $\sqrt{2}$ is the distance between two orthogonal unit vectors, and since the pooled features are non-negative, no two representations can lie farther apart than that. For $k = 1$, $S  = 0$ means that the input coincides with a training input in representation space, the pure accommodation extreme. $S  = 1$ means that it is orthogonal to all of them, the pure prediction extreme. Both are idealized limits that no real output reaches. $S$ is thus to be read comparatively, as an index of how far (close) an output lies from (to) pure accommodation (prediction). It is a surrogate for the share of an input’s real patterns on which the model was selected.

The number of neighbors $k$ sets the grain at which familiarity is measured. For $k = 1$, $S$ asks whether the training set contains an input exactly like this one (i.e., a twin). For large $k$, $S$ asks how far the input lies from a broad region of the training data, which approaches a population-level estimate. Since our criterion reads `selected for' at the level of patterns, not of individual inputs, the large-$k$ end is the more natural fit. 

As an illustration, we now apply the definition of prediction score to $M\textsubscript{RUMC}$. The upper half of Figure~\ref{Figure-2}b shows the distribution of $S $ over the correctly classified CWZ patches for $k = 1, 10, 100, 1000$. These values of $k$ correspond to fractions of the training set ranging from $0.004\%$ ($k = 1$) to $4.4\%$ ($k = 1000$). We observe that the distribution is concentrated at low values of $S$: at $k = 10$ the median of $S $ is $0.11$, nine in ten correct outputs have $S$ below $0.21$, and the largest value is $0.36$. Even at $k = 1000$, where $S$ is bound to be larger, the median is $0.19$.\footnote{Although the prediction score is defined for outputs on OOD inputs, ID inputs provide a useful reference. We obtain that median prediction score of the correct outputs on held-out RUMC patches (ID) ranges from 0.10 at $k=1$ to 0.18 at $k=1000$. These values are very close to those obtained on the CWZ patches (OOD), i.e., 0.10 and 0.19, further supporting the accommodation character of the latter.}

These outputs carry something of prediction, because their inputs instantiate patterns on which $M\textsubscript{RUMC}$ was not selected, but most of what they rest on was selected for, and in this model the accommodation part is the dominant one. That is the sense in which we suggest that most successful outputs of ML models lean strongly towards accommodation.

\paragraph{Where prediction-like outputs fail.} Which outputs, then, are more prediction-like? The lower half of Figure~\ref{Figure-2}b shows $S $ for the incorrectly classified CWZ patches. At every $k$ they lie to the right of the correct ones, with a median that is roughly twice as large (at, e.g., $k = 10$, $0.19$ vs.\ $0.11$). The median of the prediction score on incorrect outputs sits almost where the 90th percentile of that of correct ones does ($0.19$ against $0.21$ at $k = 10$, and likewise at the other values of $k$). Incorrect outputs show that $S$ is the largest where $M\textsubscript{RUMC}$ tends to fail. This is further supported by our results presented in Figure~\ref{Figure-2}c, where we display the model's own confidence, i.e., the probability it assigns to the class it picks, as a function of the prediction score. Among correct outputs, confidence is essentially $1$ only up to $S \approx 0.15$ and then falls approximately linearly with $S$. 

We further quantify the relation between the model's confidence and the prediction score through  Spearman’s rank correlation coefficient, $\rho$. This ranges from $\rho = -1$ for perfect anti-correlation to $\rho = 1$ for perfect correlation.  Between $S$ and the model’s confidence, we obtain  $\rho = -0.85$ (at $k$ = 100, and similar for other values of $k$), which signals a very strong anti-correlation. That is, the higher the prediction score of a correct output, the lower, on the whole, the model’s confidence in it. This suggests that inputs far from the training data are those on which $M\textsubscript{RUMC}$ is least certain, even when it is right.

Taken together, outputs with strong prediction character, those resulting from inputs far from anything the model selection constrained, are the ones most likely to be wrong, and the ones on which the model is least sure. This is what one would expect of a model whose success is underwritten by the patterns it was selected on: where those patterns recur, it succeeds, and where they do not, much less. If so, the successes that predictivists prize are exactly the ones that ML models are least equipped to deliver. Such successes are unlikely, and when they occur they are hard to tell apart from sheer luck.

\paragraph{Beyond a toy model.} One may object that $M\textsubscript{RUMC}$, though suggestive, is just a toy model. Surely our small convolutional network, trained from scratch on data from a single hospital, is not very representative of the ML models on which the celebrated achievements of ML-driven science rest. Granted. Yet nothing in our account depends on the nature, complexity, or scale of the model. Therefore, the picture does not change for the sophisticated models that make the headlines: even for them,  most outputs carry a strong accommodation character, too. To show this, we turn to structural biology, arguably one of the sciences in which ML has delivered its most striking successes. We consider ESM-2, a protein language model with 650 million parameters, developed by \cite{lin:2023} at Meta. It treats amino-acid sequences much like a language model treats text.
 
 ESM-2 was originally trained on UniRef50, a public database of tens of millions of protein sequences clustered at 50\% identity, by recovering residues that had been masked out. Its training set dates from 2021 or earlier. As unseen inputs, we take proteins of 50 to 500 residues, drawn at random from the manually curated Swiss-Prot entries created in 2026. These entries came into existence five years after the  ESM-2's training cutoff and cannot thus have been among the entries that selected ESM-2. For each of the 300 proteins, we hide 15\% of the residues at random and present the masked sequences to ESM-2, which fills in the gaps.\footnote{As we have mentioned above, unseen is not the same as OOD. A protein that is new to the database is not necessarily new to the model, since most new entries are relatives of old ones, and for realistic models and training sets the line between ID and OOD is blurred. Note also that curated entries are better studied than the typical sequence of UniRef50, most of which is unreviewed.} At each hidden position, the output of ESM-2 is the residue to which it assigns the highest probability. The output is correct if this residue is the one that was hidden, as recorded in the original sequence, which supplies the model-independent standard of correctness. For example, the input for a 52-residue protein is \texttt{\seqsplit{MP\_HGSLTKAG\_VRSQTPKIQPKEKHKE\_P\_V\_N\_KE\_EK\_VVKARQQAPAR}}, where each letter denotes one of the twenty amino acids, and the underscore denotes a hidden residue. At each underscore, the output of ESM-2 is the residue to which it assigns the highest probability. The output is correct if this residue is the one found at the same position in the original sequence (i.e., \texttt{\seqsplit{MPSHGSLTKAGKVRSQTPKIQPKEKHKEVPRVRNRKEYEKRVVKARQQAPAR}} in this case), and incorrect otherwise.

The original training set of ESM-2 is too large to embed. To compute the prediction score, we use in its place a proxy: a random sample of 10,000 sequences from UniRef50, restricted to entries created on or before  2021 and to sequences of 50 to 500 residues, each masked once, hiding 15\% of its residues at random, in the same way as the unseen proteins. To a good approximation, the proxy is a subset of the actual training set. Adding points to the reference can only bring the nearest neighbors of an input closer, so the prediction scores we obtain are upper bounds of those that the full training set would give. The representation of a protein is the average, over its residues, of the last hidden layer of ESM-2 applied to the masked sequence, normalized to unit length. We compute the prediction scores for $k = 1, 10, 100$ and $1000$, which correspond to 0.01\% to 10\% of the proxy training set. For the 5,583 correct outputs, the median of the prediction score is 0.16, 0.17, 0.19 and 0.22, respectively, with nine in ten outputs scoring below 0.20, 0.23, 0.26 and 0.29. For $k=10$, for instance, the prediction score ranges from a minimum of 0.09 to a maximum value of 0.43. The prediction score thus lies in the lower part of the scale, so that the proteins neither come close to coinciding with a training sequence ($S=0$) nor lie anywhere near orthogonal to all of them ($S=1$).  Even for a large model in actual scientific use, most unseen inputs lie close to what it was selected on, and, on our account, the resulting outputs carry a predominant accommodation character.\footnote{We explicitly verified that our conclusions do not qualitatively depend on the size of the proxy. Recomputing the score at $k=10$ with nested subsets of 500, 1,000, 2,000, 5,000 and 10,000 sequences, the median of the prediction score is 0.20, 0.19, 0.19, 0.18 and 0.17, i.e., it slightly decreases, as expected, as the proxy training set approaches the full training set.}

\section{Confirmation, Novelty, and Realism --- Sans Predictions}
\label{Section-5}

We conclude by sketching three vignettes to show how our account can inform various debates across the philosophy of science, touching upon confirmation, novelty, and scientific realism.

\paragraph{Confirmation sans predictions.} Predictivists of various stripes maintain that predictions are relevant, if not essential, to confirmation. Many subscribe to a comparative view, according to which predicted data confirm more strongly than the same data accommodated. We have argued that ML models never deliver pure predictions, with most of their successful outputs leaning towards accommodation. If this is accepted, a dilemma arises. Either one preserves the universal applicability of predictivism, concluding that ML models can enjoy only the weak confirmation that accommodation lends; or one rejects its universality and holds that, unlike scientific theories, ML models can be strongly confirmed in ways that do not run through pure predictions.\footnote{The dilemma bites hardest for global predictivism, according to which predictions are invariably superior to accommodations, and for strong predictivism, which holds the superiority to be intrinsic. Local predictivism escapes it by restricting that superiority to certain cases, and weak predictivism by locating the value of prediction in the features of which predictive success is a symptom.}

We take the second horn. In practice, ML models earn our confidence by operating effectively where they were not trained to operate. The further the input lies from the representations of the training data (i.e., OOD), the more the resulting correct output counts as a success, even though it still carries a predominantly accommodation character. In other words, what confirms an ML model, we argue, is neither prediction nor accommodation \textit{per se}. Rather, it is the extent to which the model captures the real patterns that underlie its empirical target. This is probed by shifting the distribution of the inputs, which raises our confidence that the model has latched onto the invariant real patterns of interest whenever the resulting outputs remain correct.

As an example, consider again $M\textsubscript{RUMC}$. A correct output on a CWZ patch is a far cry from a pure prediction, as quantified by its low yet non-vanishing prediction score, because that patch shares $\RPsh$ with the set of patches on which the model was selected. But it is informative all the same. The patch also instantiates patterns of CWZ's own, $\RPspec{CWZ}$, on which the model was \emph{not} selected, and the model gets it right in spite of them. This is evidence that the model does not rest on a single hospital staining method alone, but on features that the two hospitals share, which we take to be those underlying metastases. Correct outputs on ID inputs prove no such thing, since a model relying on RUMC's preparation procedures would do just as well on them.\footnote{The line between ID and OOD is blurred in practice, and the confirmation it supports blurs with it. Moreover, for many popular ML models, such as AlphaFold, the training set is not publicly known, so that users cannot ascertain how the model was constructed, which makes criteria for confirmation based on the construction of an ML model on their performance on shifted distributions hard to apply in practice. This is analogous to the challenge for predictivism identified by \cite{Dellsen:2025} in the context of scientific theories, according to whom a   ``theory evaluator might not know whether the data was accommodated rather than predicted in a given case. In such a case, the theory evaluator would have no grounds for placing more or less confidence in the hypothesis based on the likely presence of the relevant epistemic feature'' (p.\ 4)}

\paragraph{Novelty sans predictions.} That ML models deliver no pure predictions does not amount to saying that they deliver no novelty either. This divorce between prediction and novelty sits uneasily with much of the literature, where `novel prediction' has become a stock phrase.

What kind of novelty can ML models deliver, then? Taken literally, the temporal and heuristic criteria may count many of the ML outputs as novel. For instance, the CWZ labels were not known to us (of course, they were known to the pathologists who annotated them) when we trained $M\textsubscript{RUMC}$, and none of its patches was used to construct it. Such novelty criteria do not capture what is genuinely novel about ML models. Novelty, we suggest, resides in the mapping of the input space to the output space that the ML model has found by way of the real patterns identified in the data. While technically it is, at FSL,  a probabilistic mapping from images to labels, it can be interpreted as a mapping from the look of a patch to the presence of a metastasis. That mapping owes nothing to the history of any one input or the novelty of any one output. Above all, the mapping is obtained largely independently of the theories, rules, and laws pertaining to the relevant domain knowledge.\footnote{On the independence of ML model-building from domain knowledge see, for instance, \cite{Rattietal:2026}.} The more a mapping is found by optimization on data, and the less it is derived by reasoning from background knowledge, the more novel it is. This view is reminiscent of the notion of novelty in \citet{4988f476-2c2c-3b60-8228-802d694046c2}'s background-relative sense, according to which something counts as novel if that something is not already provided by one or more theories in the domain knowledge. 

We distinguish three kinds of novelty that ML models can deliver. They are not mutually exclusive, and we list them in decreasing order of strength. The first, \emph{novelty by supply}, arises when the domain knowledge has no theory of the connection between inputs and outputs, and the model supplies a mapping between input space and output space, found through the real patterns in the data that was hitherto unknown. This is the strongest case, since the background knowledge lacks the connection altogether. The second, \emph{novelty by rivalry}, arises when theories exist, yet the model offers a rival mapping through the identification of real patterns that differ from the ones on which the existent theory relies. This is the intermediate case, as the model does not fill a gap in the background knowledge but competes with it. The third, \emph{novelty by improvement}, arises when a theory exists and does connect inputs to outputs, but less reliably than the model does. This is the weakest case, as the ML model adds nothing in kind, only in degree.\footnote{These reflect, \emph{mutatis mutandis}, \citet{4988f476-2c2c-3b60-8228-802d694046c2}'s view.}

A given model may occupy more than one of these positions at once. Models such as $M\textsubscript{RUMC}$ illustrate this by combining novelty by rivalry with novelty by improvement. Pathologists surely have diagnostic criteria that connect a patch to a verdict, yet such criteria do not involve the examination of pixels, or at least  not of pixels \textit{qua} pixels. The model, by contrast, works from the pixels of a stained patch, and the real patterns through which it separates metastatic from non-metastatic tissue need not be those that pathologists rely on. Moreover, models of this kind could provide novelty by improvement, in that they could lead to a more accurate diagnosis. This could be assessed by comparing the $M\textsubscript{RUMC}$'s performance on CWZ patches with that of pathologists at RUMC.

\paragraph{No-miracle argument sans predictions.} Our account directly impacts recent debates on scientific realism. In the case of scientific theories, scientific realism is often defended on the basis of \cite{Putnam:1975}'s No-Miracle Argument (NMA), according to which, were it not for their approximate truth, the empirical success of  theories would be left without explanation, akin to a miracle. More recently, \cite{RN687} and \cite{RN683} have offered a reformulation of the NMA that is applicable to ML models. Let $A$ be an arbitrary model in the domain of predictive ML models in science. The argument takes the following probabilistic form:

\begin{quote}
\begin{tabular}{@{}l@{\quad}l@{}}
  (a) & $P(S_A \mid {\approx}A \wedge F)$ is high. \\
  (b) & $P(S_A \mid \neg{\approx}A \wedge F)$ is low. \\ \cline{2-2}
  (c) & {Therefore,} $P({\approx}A \mid S_A \wedge F)$ is high.
\end{tabular}
\end{quote}
Here, $S_A$ stands for `$A$ is empirically successful', defined on the basis of the number of novel predictions scored by the ML model; $\approx A$ stands for `$A$ quasi-represents its task-related empirical targets', where `quasi-' denotes the representational force of the medium towards its target, as described by \citet{254c1812-f25d-37fd-a062-eaf9c32dc5a1}, without the `aboutness' usually ascribed to representations; and $F$ is the relative frequency of success.

We understand the notion of `quasi-representation' of \cite{RN683} to be close to what we call real patterns, that is, a compressed description that captures the regularities in the data. If so, the argument licenses the inference from the empirical success of ML models to the conclusion that they latch onto real patterns of their intended target. But what is `empirical success'? According to \cite{RN687}, it is the ability of ML models to produce novel predictions. On our account, however, ML models do not entail pure predictions. If `novel prediction' is read strictly, $S_A$ is then almost never satisfied, to the effect that premise (a) is false and the argument is unsound.

One may consider a weaker reading, on which the notion of novel prediction is taken to mean the production of correct outputs on unseen inputs. \citet{RN683} seem to entertain this idea, when they ``proceed on the assumption that a DLM [Deep Learning Model] cannot achieve `empirical success' merely by achieving predictive success concerning items in its training set'' (p.\ 6). The novelty at stake would then be that of unseen versus seen inputs. However, an unseen input can lie very close, in distribution, to the training data, and what the model captures at FSL includes the regularities of the measurement apparatus along with those of the intended target, as illustrated by $M\textsubscript{RUMC}$. Given the opacity of ML models, it is hard to single out the quasi-representation of the target from whatever else the model has captured, so that success on unseen inputs is not evidence that the model has quasi-represented it.

This can be remedied by considering significant distribution shifts, as we have shown by running $M\textsubscript{RUMC}$ on CWZ patches.  If the ML model remains correct when the apparatus-specific patterns change, our confidence that it has captured the real patterns of the intended target rises. If the NMA is to apply to ML models, then, it had better rest on success under such shifts, that is, on OOD outputs. Yet these are still, for the most part, accommodation-like. The realist would thus ground their commitment on accommodations, which is uncomfortable: the NMA draws its force from success that would be miraculous if the model had not captured the target, and an accommodated success is no such miracle, since model selection explains it. Proponents of the NMA in ML thus face a challenge. If they retain their own conception of empirical success as `novel prediction,' their argument is unsound, since the premises cannot be met. If they relax it to include success on OOD inputs, they ground a realist commitment on outputs that are accommodation-like, and the miracle the argument relies on is gone.

\bibliographystyle{apalike}
\bibliography{Bibliography}
\end{document}